# A long term study of Mars mesospheric clouds seen at twilight based on Mars Express VMC images.

J. Hernández-Bernal[1,2] , A. Sánchez-Lavega[1] , T. del Río-Gaztelurrutia[1] , R. Hueso[1] , E. Ravanis[3,4], A. Cardesín-Moinelo[3,5] , S. Wood[6], D. Titov[7]

[1]Dpto. Física Aplicada I, EIB, Universidad País Vasco UPV/EHU, Bilbao, Spain [2]Aula EspaZio Gela, Escuela de Ingeniería de Bilbao, Universidad del País Vasco UPV/EHU, Bilbao, Spain [3]European Space Agency, ESAC, Madrid, Spain [4] Department of Earth Sciences, University of Hawai'i, Honolulu, HI, USA [5]Instituto de Astrofísica e Ciências do Espaço, Obs. Astronomico de Lisboa, Portugal [6]European Space Agency, ESOC, Darmstadt, Germany [7]European Space Agency, ESTEC, Noordwijk, The Netherlands

Corresponding author: Jorge Hernández-Bernal (jorge.hernandez@ehu.eus)

> **Key Points:**
> - We present a new methodology to detect clouds at twilight and measure their altitude. We find 407 cases, some at altitudes over 80km.
> - High altitude clouds appear most often in mid-latitudes during the local winter. This is a new trend when compared to previous studies.
> - High altitude clouds concentrate aerographically in a southern belt that includes Terra Cimmeria, and in clusters on northern planitias.

## Abstract

We present the first systematic study of clouds observed during twilight on Mars. We analyze images obtained by the Visual Monitoring Camera (VMC) on Mars Express between 2007 and 2020. Using an automated retrieval algorithm we found 407 cases of clouds observed at twilight, in which the geometry of the observations allows to derive the minimum altitude, revealing that many of these clouds are in the mesosphere (above 40km and up to 90km). The majority of these mesospheric clouds were detected in mid-latitudes at local autumn and winter, a new trend only hinted at by previous studies. In particular, we find a massive concentration of clouds in the southern mid-latitudes between Terra Cimmeria and Aonia, a region where high altitude events have been previously observed. We propose that there is an unknown mechanism in these regions that enhances the probability to host high altitude clouds around the southern winter solstice.

## Plain Language Summary

During twilight, when the sun is below the horizon, its light can still reach clouds or mountains high above the surface, making them bright features on the dark background. This effect is sometimes seen on noctilucent clouds on Earth, and also in the mountains on the Moon. On Mars, it was first observed by ground based observers in the 1890s, and occasional observations have been later reported. We present here the first systematic study of such clouds on Mars as observed from space by the Visual Monitoring Camera (VMC, also known as the Mars webcam) onboard Mars Express. The study of clouds at twilight reveals information about their altitude, and the state of the atmosphere at this moment of the daily cycle. We analyze the occurrence of these clouds and find some new trends that previous observations had only hinted at.

## 1. Introduction

During twilight, when the sun is below the horizon, high altitude structures like mountains or clouds can receive the light of the sun, and appear bright over the dark of the night. On Earth, we can see this effect from the ground in high altitude noctilucent clouds (Gadsden & Schröder, 1989). On Mars, bright patches at the terminator were first observed by telescopic observers in the 1890s (Holden et al., 1890; 1894; Campbell, 1894; Douglass, 1895; 1897), but due to the lack of topographic models at the epoch it was not clear whether they were due to topography or atmospheric aerosols. Today, current topographic data (Smith et al., 2001) allow us to deduce that most of these observations of bright features in the night side of the terminator of Mars are atmospheric aerosols (water ice, CO2 ice, and dust, see e.g. Zurek 2017 or Clancy et al., 2017). For simplicity, we refer in this paper to all the aerosol patches as "clouds", no matter whether they are made of condensates or dust, as we cannot determine the composition of aerosols from our observations.

Clouds at twilight on Mars have been detected by the Hubble Space Telescope (HST; James et al., 1996), the Thermal Emission Imaging System (THEMIS; Inada et al. 2007; McConnochie et al. 2010) onboard Mars Odyssey, and the Imaging Ultraviolet Spectrograph (IUVS; Connour et al., 2020) on board the MAVEN (Mars Atmosphere and Volatile Evolution) mission. However, until 2014 most orbiters on Mars were placed in afternoon sun-synchronous orbits, and thus, few orbital observations of regions in twilight are available (Hernández-Bernal et al., 2020). Among the surface missions, Viking landers performed observations of the brightness of the twilight, but did not find clouds (Kahn et al., 1981), and the Pathfinder lander found a cloud 60-100km over the surface early before sunrise (Smith et al., 1997; Clancy & Sandor 1998). More recently, the Curiosity rover has also imaged clouds at twilight (Newman, 2019).

In this paper, we study twilight clouds (Figure 1) with the aid of the Visual Monitoring Camera (VMC; Sánchez-Lavega et al., 2018; see also Ravanis et al., 2020) onboard Mars Express (MEX). VMC is probably the best suited among available instruments for this kind of study, as it has a wide field of view, takes regular full disk images showing the terminator of Mars, and its archive covers 6 Martian Years (MY; from MY 29 to MY 35).

While different types of twilight are clearly defined for the case of Earth (e.g. Seidelmann et al., 1992), no formal definitions exist for the twilight on Mars. In this paper, we consider for statistical purposes the region from the terminator to 15° into the night (incidence angle 90°-105°) as twilight. This distance from the terminator corresponds to a minimum altitude of 120 km (see section 2, and supporting text S2) and while high altitude events might overtake this altitude (Sánchez-Lavega et al., 2015), we did not find such high altitude events in this work.

The Martian Climate is characterized by a "cloudy season" (Solar Longitude, Ls ~0-180°), and a "dusty season" (Ls ~135°-360°). Dust storms of different scales predominate in the dusty season (Kahre et al., 2017) and water ice clouds are most common in the "cloudy season", but both water clouds and dust storms can be found during the whole year in different locations. Most water ice clouds can be classified in two main regimes (Clancy et al., 2017; see also Fig 2 in Wang & Ingersoll, 2002): (1) the Aphelion Cloud Belt (ACB; Clancy et al., 1996), which dominates the cloudy season; (2) and the Polar Hoods (PH; Benson et al, 2010; 2011), which appear around the polar regions mostly during the local autumn and winter.

Clouds bright at twilight are necessarily high above the surface. Following the discoveries by Clancy et al. (2007), Montmessin et al. (2006a; 2006b; 2007), Vincendon et al. (2011) and others, Martian mesospheric clouds (40km and above) have been widely investigated in recent years. They are typically found at maximum altitudes of 80-90km, and are more common in equatorial regions in the cloudy season, and in midlatitudes in the dusty season. They can be made of dust, water ice, or $CO_2$ ice, and the climatology of different compounds is not always clear from observational studies, due to observational limitations. Based on afternoon and mid-night systematic observations from MRO, Clancy et al. (2019) find that dust is rare in the mesosphere, $CO_2$ aerosols are more common in equatorial regions at Ls 0-160°, and $H_2O$ aerosols prevail at Ls 160°-360°, in the latitude range 50°N-50°S, with a minimum at low latitudes, in agreement with the trends found by Sefton-Nash et al. (2013). The formation of $CO_2$ clouds, which seems to be common in the mesosphere, is still challenging for models. $CO_2$ ice requires very low temperatures to condense, and thus the diurnal cycle is a relevant factor for their study (González-Galindo et al., 2011). There have been some reports of mesospheric clouds around twilight, in the evening by THEMIS (Inada et al., 2007; McConnochie el al., 2010), and in the early morning by MEX/OMEGA and MEX/HRSC (Montmessin et al., 2007, and Määttänen et al., 2010), often in mid-latitudes during the local autumn. For reviews on mesospheric clouds see Määttänen et al. (2013) and Clancy et al. (2017)

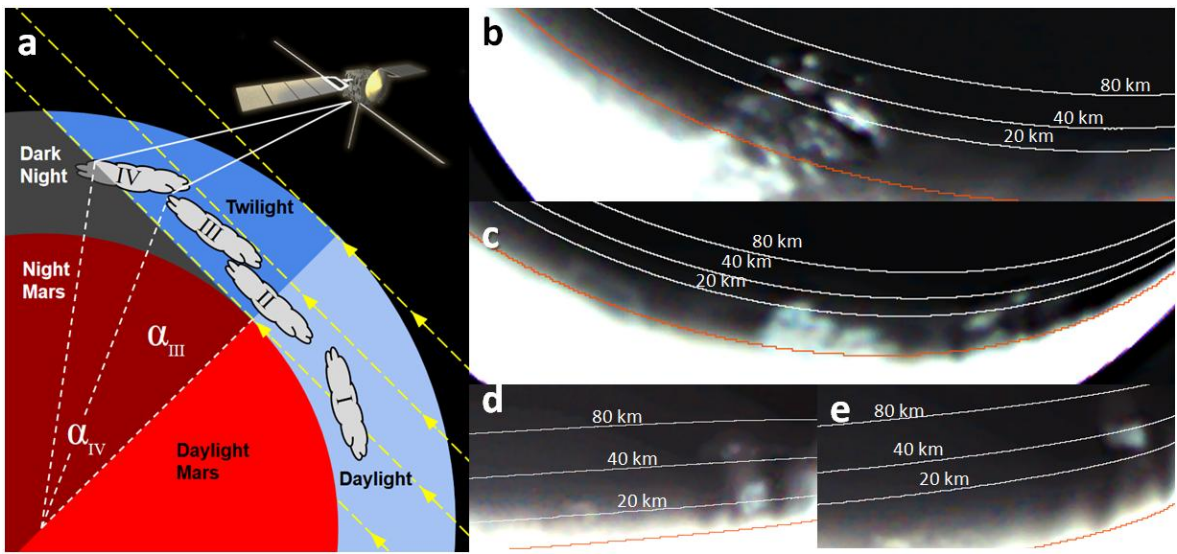

**Figure 1.** Scheme of cloud detection from orbit (left), and cloud examples from VMC images (right). (a) The daylight, twilight, and dark night regions of the atmosphere are indicated. Clouds III and IV are visible from space as isolated bright patches over the dark of the night, while II is seen as a bright patch connected to the terminator. Note that α is the incidence angle minus 90°; (b) A cluster of clouds over Terra Cimmeria on Ls 123° (MEXVMC_1800430001, 2018-02-01); (c) Some discrete clouds (right) and a cloud connected to the terminator (left) around Alba Patera at Ls 343° (MEXVMC_1700560001, 2017-04-04); (d,e) A recurrent cloud observed on 2020-03-08 and 2020-03-10 at Ls 164°, around Terra Sirenum (MEXVMC_2000970003 and MEXVMC_2001010003). In (b-e) minimum-altitude curves are drawn in gray and the terminator in red.

The systematic study of aerosols in the twilight presented here provides information about the state of the atmosphere at different altitudes in different locations and seasons of Mars. In section 2, we explain our dataset and methods, in section 3, we show our results, and in section 4 we discuss our results and conclusions.

## 2. Dataset and Methods

We base our study on the MEX/VMC image dataset (Ravanis et al., 2020), which includes more than 2000 independent observations (each observation lasts several minutes and contains a number of consecutive images) in MYs 29-35, mostly from Mars Express apocenters, where VMC captures full disk images of the planet.

A first estimation of the minimum altitude necessary for the clouds at twilight to be illuminated, deduced from simple geometry (Fig. 1a), is: $h_{min} = R_{Mars} \tan(\alpha) \tan(\alpha/2)$. This corresponds to actual altitude for clouds that are at the limit of the twilight (cloud example IV in Fig. 1a). However, this simple calculation of minimum altitudes ignores the non-spherical shape of the planet and the parallax of the observations. To include these effects we have developed a numerical method, detailed in the supporting material text S2. For image navigation, geometry, topography, and Mars time calculations, we used the methods described in Hernández-Bernal et al. (2020).

We performed an automated search of bright patches in the dark (night) side of the terminator considering only isolated patches (i.e. bright areas disconnected from the terminator) at least 20km over the areoid. Therefore, we only detect discrete clouds (e.g. cloud examples III and IV in Fig. 1a), and bright hazes crossing the terminator into the dark side of Mars are ignored (e.g. cloud example II in Fig. 1a). In our automated search, MEX/VMC images showing the dark (night) side of Mars with appropriate exposure are first selected. Then, the selected images are analyzed in search of bright features in the dark side of the planet (clouds). The results from the automated analysis are supervised visually to discard false positives. Finally, the parameters for each cloud, including minimum altitude, location, area, date, Ls, and MY, are stored. The details of our analysis pipeline are presented in the supporting material text S1. The low quality RGB Bayer pattern of VMC color images (Ormston et al., 2011) is not sufficient to determine the composition of these clouds.

## 3. Results

We have analyzed 4621 images in 1891 independent observations spanning from MY28, Ls 344º, to MY35, Ls 229º (November 2007 to June 2020). The coverage trough time is not homogeneous (Ravanis et al., 2020), and it is especially low in MYs 31 and 32. Due to the MEX orbit, most of the observed areas correspond to morning twilight (see supporting Fig. S1 for local time bias). In total, we have detected 407 clouds.

## 3.1. Altitude of the clouds

In Figure 2, we show the statistical distribution of clouds at twilight according to their minimum altitude, and we provide some examples of the highest detected features. Typically, the uncertainty of our calculated minimum altitudes is below 5-10km (the uncertainty is higher for higher altitudes). The highest clouds detected have a minimum altitude of around 80 km over the areoid, which is in agreement with previous studies on high altitude aerosols (Määttänen et al., 2013).

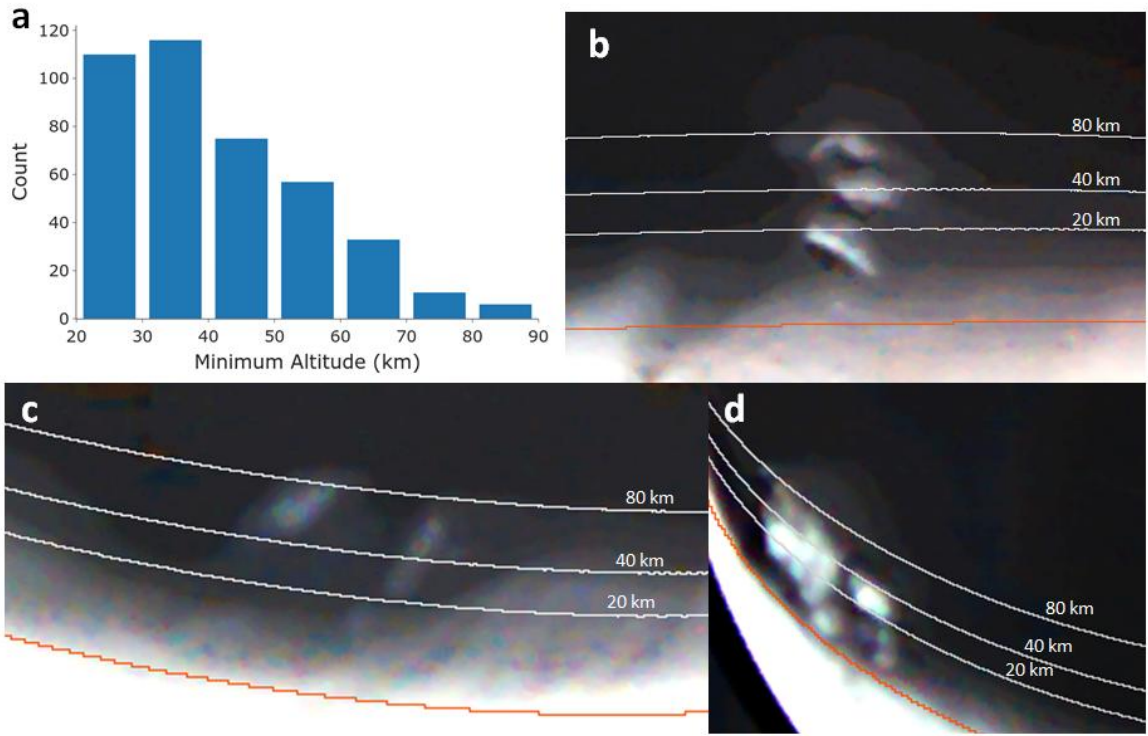

**Figure 2.** Minimum altitude of detected clouds. (a) Statistical distribution of the detected minimum altitudes. (b) Cloud with altitude ~81 km over Terra Cimmeria (MEXVMC_0800560021, 2008-09-07). (c) Cloud ~82 km at 80°N  (MEXVMC_0900110028, 2009-04-09). (d) Cloud ~89 km over Argyre Planitia (MEXVMC_1904350006, 2019-11-25).

## 3.2. Areographic and Seasonal distribution

The areographic distribution of clouds at twilight shows that they are more common in specific areas, grouped in three latitudes (Figure 3). The southern group forms a prominent belt around 45°S, and most clouds there seem to concentrate in three regions distributed in the longitude range 120°-360°E (Terra Cimmeria, Terra Sirenum, and Aonia Terra). Other concentrations appear to the west of Noachis Terra and on the southern border of Hellas Planitia. In the equatorial group, most clouds are around the Tharsis area and also in the regions of Margaritifer Terra and Tyrrhena Terra. It is worth noticing that we do not find clouds higher than 60km in equatorial latitudes, and most detected clouds in this region are below 40km. Fewer clouds are found in the northern latitudes, around 45ºN, mostly in the regions of Acidalia and Arcadia Planitia. In general clouds are more common in the longitude range 180º-360ºE at all latitudes, and in the case of the southern belt, this limit shifts westward to 130ºE including Terra Cimmeria.

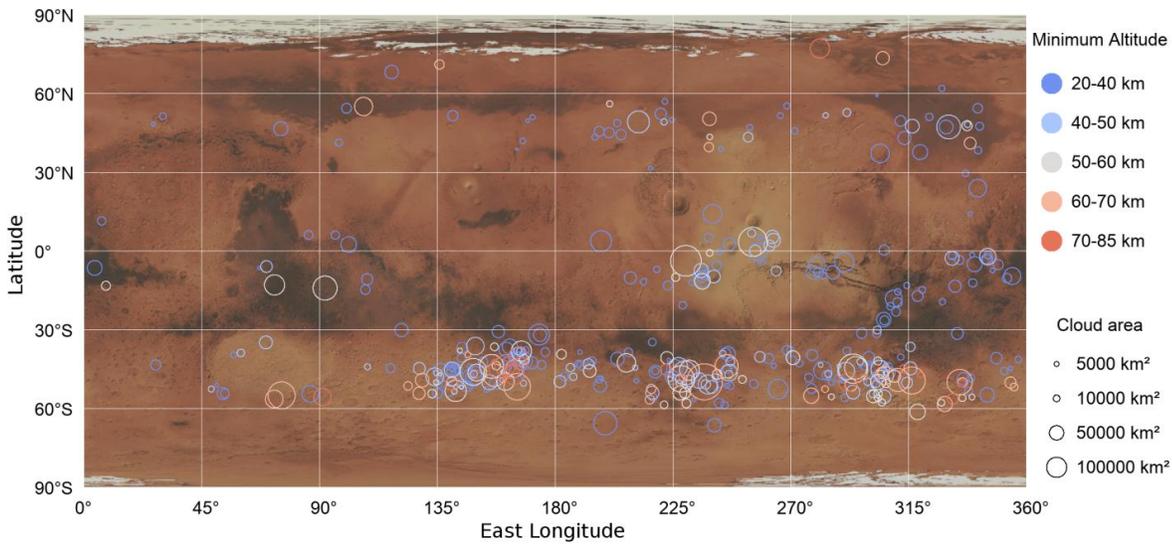

**Figure 3.** Areographic distribution of detected clouds with minimum altitude above 20 km. Clouds are indicated with circles. Color code indicates minimum altitude and size of the circle indicates area. The coverage of the different areas can be found in supporting material Figure S1.

In Figure 4 we show the seasonal distribution of clouds from observations in different years, including morning and evening twilight. The plot shows that clouds at twilight are more common in the cloudy season (Ls 0º-180º). The southern belt shows an intense activity in the Ls range 40-180º. Clouds from the equator are observed in the Ls range 45º-140º whereas a northern cloud concentration at mid-latitudes takes place in the Ls range 330º-20º, at the end of the dusty season. We detected several clouds at lower minimum altitudes on latitudes 0º-60ºS in Ls 225º-290º, which might correspond to dust activity, since this range of Latitude and Ls was only observed in MY34, which was characterized by a global dust storm in Ls 185º-280º (Sánchez-Lavega et al., 2019; Montabone et al., 2020). In the supporting material, we separate the results for the morning (Fig. S3) and evening (Fig. S4) twilight. We find that the southern belt is present at both morning and evening. However, the sampling of the evening is poor in other regions, preventing us from extracting further conclusions on LTST dependence.

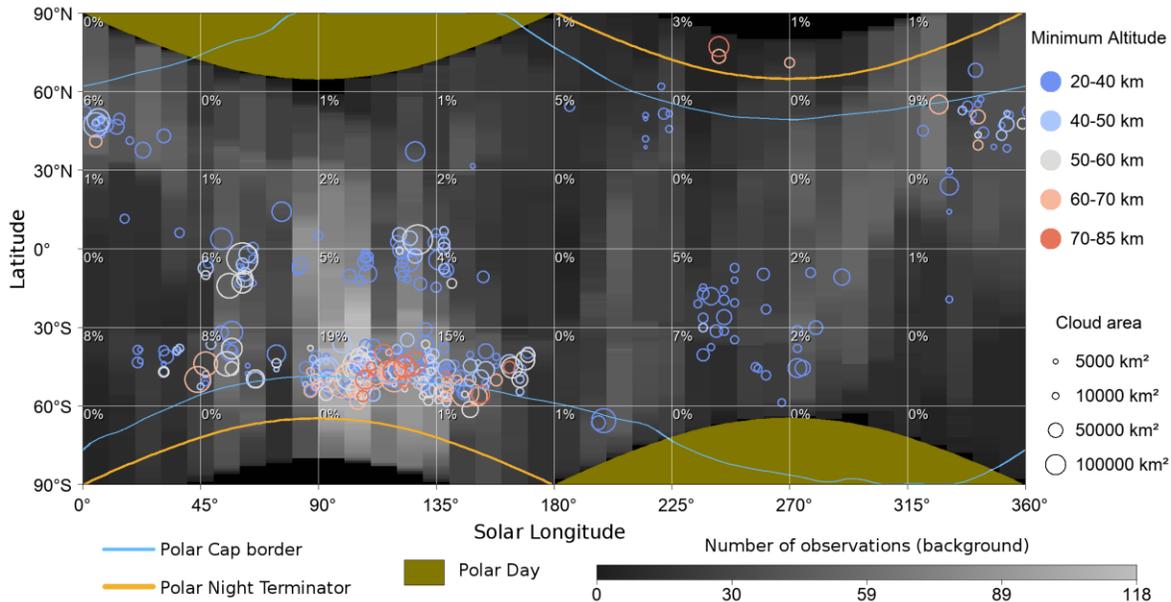

**Figure 4.** Seasonal distribution of detected clouds with minimum altitude over 20 km and from MYs 29-35. For each rectangle of the grid, the percentage of positive observations (containing at least one cloud over 20 km) is indicated in the corner. The border of the polar caps was plotted from data presented in Piqueux et al. (2015).

The north polar night was only observed in MY29, between Ls 240° and 270°. Three high clouds were detected (two of them shown in Fig. 2c), with minimum altitudes of 60-80km.

## 4. Discussion and Conclusions

Aerosols observed in twilight are distributed mainly in three groups: one over the equator (Ls 40°-140°), one at northern mid-latitudes around 45°N at the end of the local winter (Ls 330°-30°), and the southern belt around 45°S during the local autumn and winter (Ls 40°-160°). A considerable fraction of the detected clouds are at minimum altitudes of 40 km or more, and thus we consider them to be in the mesosphere. When the minimum altitude is lower, clouds could be located in the troposphere., where the ACB and the PHs dominate cloud trends.

The equatorial group could be related to the ACB, which takes place in the same latitude range and season. The detected clouds in equatorial latitudes are not very high (most cloud tops are below 40km), which is in agreement with the altitudes reported by McCleese et al. (2010) for the ACB. The mid-latitude groups might have some relation with the PHs, that take place in autumn and winter. Our southern twilight belt appears at the same time as the South Polar Hood (SPH). Benson et al. (2010) reported top altitudes of 30-40km for the SPH, but we find a considerable number of clouds in the mesosphere, higher than 40 km, in the Ls range 90°-150°. Our northern group only starts at the end of the northern winter (Ls >330°). Again, Benson et al. (2011) reported top altitudes of ~40 km for the NPH, and we find a number of clouds at higher altitudes.

Our study reveals that mesospheric clouds (with minimum altitudes of at least 40 km) are common in the southern belt during the local winter. Previous studies, which occasionally reported clouds at mid-latitudes during the local autumn around twilight (Montmessin et al., 2007; Määttänen et al. 2010; Scholten et al., 2010; McConnochie et al., 2010; Sánchez-Lavega et al., 2018, see also Fig. 6 in the review by Määttänen et al., 2013), constitute hints of the trend we report here. In the case of the northern belt, we only detect high clouds after Ls 330°, but this result must be taken with caution, since the sampling of the region is poorer, particularly earlier in the season. McConnochie et al. (2010) reported the presence of high clouds at northern mid-latitudes in an earlier Ls range 200°-300° suggestive of a similar trend as in the Southern belt. It must be noted that McConnochie et al. (2010) detected clouds in the evening twilight, while our northern detections correspond to morning twilight, but this is probably not significant since our data is clearly biased towards morning twilight, as can be seen in supporting figure 2. In supporting figures S3, S4 we show respectively morning and evening twilight detections, showing the presence of the southern belt at both morning and evening twilight. In agreement with the review by Määttänen et al. (2013), we find lower mesospheric cloud activity around the aphelion (see fig. 4, around Ls 70°), this trend is also apparent in Clancy et al. (2019, their figure 10).

It is worth noting that our southern belt is very prominent, both in terms of the quantity of detected clouds, especially in the Ls range 90°-180°, when we find twilight clouds in ~19% of our observations, and because of the frequency of high altitude apparitions (Fig. 4).

We cannot determine the aerosol nature of our detections from our observations alone. However, given the seasonal distribution of these clouds, and by comparison to the studies of limb spectra by Sefton-Nash et al. (2013) and Clancy et al. (2019) from data obtained by the Mars Climate Sounder, most of these clouds are likely to be condensates, and some of them might be made of CO2. Montmessin et al, 2007 and Määttänen et al., 2010 reported two mid-latitude mesospheric clouds at latitudes 49ºS and 47ºN confirmed to be made of CO2 by OMEGA. Temperatures predicted by atmospheric models such as the LMD-MGCM are not low enough for CO2 to condense, but observed CO2 clouds occur in regions where temperatures predicted by models are closest to condensation (González-Galindo et al., 2011).Those models could indicate that CO2 condensation is possible in our southern belt (see Figure 3 in González-Galindo et al. 2011).

H2O clouds have also been observed at mesospheric altitudes (e.g. Vincendon et al. 2011; Clancy et al., 2019). While not much water vapor is expected at mesospheric altitudes in the southern latitudes during the aphelion season (Federova et al. 2020 report low concentrations, and the LMD-MCD (Forget et al., 1999; Millour et al., 2009) predicts concentrations around 1-3 ppmv), and temperatures are low enough to produce water-ice condensates. Another possibility is that the water concentration is actually enhanced due to the upward branch of the Hadley cell around the northern solstice, injecting water vapor coming from the northern latitudes (Heavens et al., 2011). The northern group occurs in the dusty season, and thus the presence of high altitude dust is expected. It is also coincident with the H2O mesospheric cloud trends reported by Clancy et al. (2019) and temperatures and water-mixing rations predicted by the LMD-MCD are consistent with water ice condensation.

Dust is not very common in the aphelion season and thus our southern belt is not likely to contain dust, but several clouds observed around Ls 225º-290º in MY35 in the southern hemisphere (Fig. 4), might correspond to dust activity during the Global Dust Storm GDS 2018 (Sánchez-Lavega et al., 2019; Montabone et al., 2020).

Regarding the distribution in longitude, our detected clouds are more common around longitudes 180º - 360º E, in good agreement with previous works (review by Määttänen et al., 2013; Aoki et al., 2018; Sánchez-Lavega et al., 2018; Clancy et al., 2019). The longitudinal distribution of twilight clouds is inhomogeneous (Fig. 3) despite a quite homogeneous sampling of each latitude (see Figure S1); they are most common on Terra Cimmeria, Terra Sirenum, and Aonia Terra. An extremely high-altitude plume was observed ⸱ 200 km over Terra Cimmeria, also in the early morning around the northern solstice (Sánchez-Lavega et al., 2015). This region seems to be a common scenario of high altitude events, especially in the early morning, as shown in Fig 7 of Pellier (2012), using HST data. We propose that the regions of Terra Cimmeria, Terra Sirenum, and Aonia Terra, might have some unknown mechanism to produce high altitude hazes around the southern winter solstice.

The observation of the twilight can provide valuable information about water vapor distribution and the temperature structure required to form condensates. Currently available orbiters suitable for this kind of study are Mars Express (e.g. Montmessin et al., 2007; Määttänen et al., 2010; this work), MAVEN (e.g. Connour et al., 2019), MOM (Arya et al. 2015), and Mars Odyssey (McConnochie et al., 2010). The new Tianwen-1 (Wan et al., 2020) and HOPE (Sharaf et al., 2020) orbiters will also be able to perform such investigations.

## Acknowledgments, Samples, and Data

This work has been supported by the Spanish project AYA2015-65041-P and PID2019-109467GB-I00 (MINECO/FEDER, UE) and Grupos Gobierno Vasco IT-1366-19. JHB was supported by ESA Contract No. 4000118461/16/ES/JD, Scientific Support for Mars Express Visual Monitoring Camera. The Aula EspaZio Gela is supported by a grant from the Diputación Foral de Bizkaia (BFA). We acknowledge support from the Faculty of the European Space Astronomy Centre (ESAC). Special thanks are due to the Mars Express Science Ground Segment and Flight Control Team at ESAC and ESOC.

## Data Availability Statement

The VMC dataset used in this investigation is available in the ESA PSA (Planetary Science Archive) at:

https://archives.esac.esa.int/psa/#!Table%20View/VMC%20(Mars%20Express)=instrument